\documentclass[prl,twocolumn,letterpaper,superscriptaddress,amsmath,amssymb,floatfix,showpacs]{revtex4}
\usepackage{graphicx}
\usepackage{epsfig}
\usepackage{color}

\newcommand{\red}[1]{#1}

\newcommand{\ignore}[1]{}

\begin{document}
\title{Coexistence of gapless excitations and commensurate
charge-density wave in the 2H-transition metal dichalcogenides}
\author{Ryan L. Barnett}
\affiliation{Department of Physics,
  Harvard University, Cambridge, MA 02138}
\author{Anatoli Polkovnikov}
\affiliation{Department of Physics,
  Harvard University, Cambridge, MA 02138}
\author{Eugene Demler}
\affiliation{Department of Physics,
  Harvard University, Cambridge, MA 02138}
\author{Wei-Guo Yin}
\affiliation{Physics Department,
 Brookhaven National Laboratory, Upton, NY 11973}
\author{Wei Ku}
\affiliation{Physics Department,
 Brookhaven National Laboratory, Upton, NY 11973}

\begin{abstract}
An unexpected feature common to 2H-transition metal dichalcogenides
(2H-TMDs) is revealed with first-principles Wannier functions
analysis of the electronic structure of the prototype 2H-TaSe2: The
low-energy Ta \red{``$5d_{z^2}$''} bands governing the physics of
charge-density wave (CDW) is dominated by hopping between
next-nearest neighbors. With this motivation we develop a minimal
effective model for the CDW formation, in which the unusual form of
the hopping leads to an approximate decoupling of the three
sublattices. In the CDW phase one sublattice remains undistorted,
leaving the bands associated with it ungapped everywhere in the
Fermi surface, resolving the long-standing puzzle of coexistence of
gapless excitations and commensurate CDW in the 2H-TMDs.
\end{abstract}
\date{\today}

\maketitle

Charge-density waves (CDWs) in solids has  been a topic of central
interest in condensed matter physics for many years \cite{Gruner94}.
Recent scanning tunneling microscopy experiments showing a periodic
modulation in the local density of states in cuprate superconductors
\cite{Howald03,Hanaguri04}, has reinvigorated such interest. Despite being one
of the earliest discovered class of materials which exhibit a CDW at
\ignore{sufficiently} low temperatures~\cite{Wilson74}, many
properties of the 2H-transition metal dichalcogenides (2H-TMDs) are
still not understood, leading to much recent theoretical
\cite{Rice75,Neto01,Uchoa04} and experimental
\cite{Yokoya01,Liu98,Straub99,Liu00,Tonjes01,Valla00,Rossnagel01,
Valla04} research effort (for a review, see
Ref.~[\onlinecite{Withers86}]).  Two key issues concerning the CDW
phase in these materials deserve the most attention.  First,
controversy exists between different experimental groups on the
driving mechanism of CDW originating from quantitative differences
between the angle-resolved photoemission spectroscopy (ARPES) data.
While some experimental results \cite{Liu98,Straub99,Liu00,Tonjes01}
suggest that the hexagonal Fermi surfaces around the $\Gamma$ point
are consistent with the CDW nesting vector, others
\cite{Valla00,Rossnagel01,Valla04} indicate that this Fermi surface
is too large to give the correct nesting vector. Second, and of a
more qualitative nature, ARPES measurements
\cite{Liu98,Straub99,Liu00,Tonjes01,Valla00,Rossnagel01, Valla04}
find no evidence of a gap opening on the hexagonal Fermi surface, in
direct contrast with traditional wisdom of CDW materials.

In this Letter, we focus on the latter issue and suggest a simple
picture for why such gapless excitations are permitted in the CDW
phase. Using density-functional theory, the electronic structure of
prototype 2H-TMD, 2H-TaSe$_2$, is analyzed with a newly developed
Wannier function \red{(WF)} approach~\red{\cite{Ku02,Grenier05}},
and a striking feature is revealed: the low-energy bands near Fermi
surfaces, which governs the physics of CDW, is dominated by hopping
between second-nearest neighbors. This special nature of hopping, in
combination with the triangular lattice vectors, effectively splits
the system into three weakly coupled triangular sublattices. Since
the CDW \red{state} gives distortion of only two of the sublattices,
whose stability is illustrated with a simple model, such unique
electronic structure naturally leaves the bands associated with the
undistorted sublattice ungapped in this CDW phase, resolving the
puzzle of the observed gapless excitations along the nested regions
of the Fermi surface.

\begin{figure}[b]
\includegraphics[width=3in]{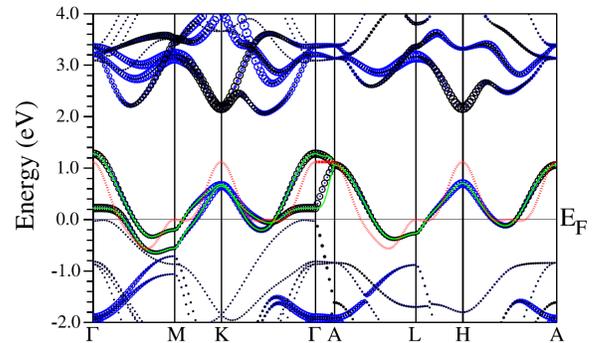}
\caption{\label{Fig:band}%
First-principles band structure (dots) with the $d_{z^2}$ (black
circles) and $d_{xy}/d_{x^2-y^2}$ (blue circles) characters shown.
The bands below $-0.7$ eV are mainly Se $p$ bands. Also shown are
the bands constructed from low-energy WFs (green solid lines) and
\red{a 2D `nesting' model (red dotted line; see text).}}
\end{figure}

The lattice structure of 2H-TMDs consists of stacked layers of
2D-triangular lattices of transition metals (e.g$.$ Ta or Nb)
sandwiched between layers of chalcogen atoms (e.g$.$ S or Se). Rough
estimation of the ionization leads to the Ta$^{4+}$Se$^{2-}$
configuration with one valence $5d_{z^2}$ electron left per Ta atom
that forms the metallic bands at the Fermi level. Indeed, the band
structure from our first-principles calculations \cite{note:wien}
(see Fig.~\ref{Fig:band}) shows a strong $d_{z^2}$ character in the
two metallic bands \red{corresponding to two weakly coupled TaSe$_2$
sandwiches per unit cell. Unexpectedly, little Se $p$ characters are
found in these two bands. The calculated low-energy bands agree well
with experiments\ignore{~\cite{Pillo00}}, except that the saddle
bands on the $\Gamma$K and AH lines are not as flat and close to the
Fermi level as reported~\cite{Liu98,Liu00}.}

\begin{figure}
\includegraphics[width=3in]{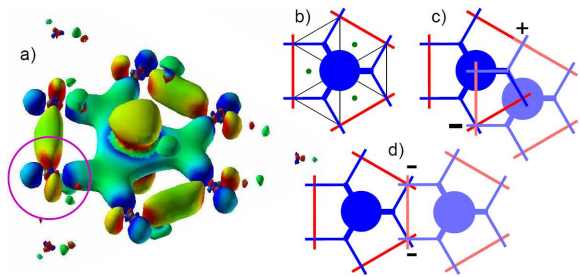}
\caption{\label{Fig:wannier} \red{a) Low-energy WF centered at Ta
sites with $a_g$ ($d_{z^2}$) symmetry, colored to show its gradient
from positive (red) to negative (blue). Notice the local
$d_{xy}/d_{x^2-y^2}$ symmetry in the hybridization tail (circled)
located at neighboring Ta sites. b) Schematics of the WF in the
layer of the Ta triangular lattice, with a similar color scheme
giving the sign of the WF and three small dots marking the positions
of Se atoms in the next layer. c) and d) Schematics of phase
interference in hopping to first and second nearest neighbors,
respectively. }}
\end{figure}

Based on the first-principles ground state, the low-energy Hilbert
space can be accurately extracted via local WFs (see
Fig.~\ref{Fig:wannier}), which we constructed by extending recently
developed energy-resolved method~\cite{Ku02,Grenier05} to
incorporate desired symmetry~\cite{Ku05}. As expected, the WF
located at each Ta site has strong $d_{z^2}$ symmetry near the
center, before extending its \emph{unusual tails of
$d_{xy}/d_{x^2-y^2}$ symmetry} to the nearest neighboring Ta sites
due to strong hybridization with the $d_{xy}/d_{x^2-y^2}$ orbitals
near the $K$ and $H$ points (see Fig.~\ref{Fig:band}). This
particular shape of the WF results in an intriguing feature in the
hopping integral (evaluated via
$t_{\mu\nu}=\left\langle{\mu\left|h^{DFT}\right|\nu}\right\rangle$
with density functional theory Hamiltonian, $h^{DFT}$, and Wannier
states $\left|\mu\right\rangle$ and $\left|\nu\right\rangle$). That
is, the second neighbor hopping, $t_2=115$ meV, overwhelms the first
neighbor hopping, $t_1=38$ meV, due to remarkable phase cancelation
in the latter case (to be discussed in more detail below). In
addition, interlayer hoppings are found to be comparable to
first-neighbor in-plane hopping with $t_{\perp,1}=29$ meV and
$t_{\perp,2}=23$ meV.

\red{A simple microscopic picture for the unexpected dominance of
second-neighbor hopping can be obtained  from the
symmetry of the WFs. As shown in Fig.~\ref{Fig:wannier}(c) and (d),
the contributions to the hopping parameters between neighboring WFs
come mainly from overlap of their hybridization tails, since the
tail-center ($d_{z^2}-d_{xy}/d_{x^2-y^2}$) overlap gives negligible
contribution due to its odd parity. While the first-neighbor hopping
suffers seriously from the phase cancelation (illustrated by the
\emph{opposite} sign in Fig.~\ref{Fig:wannier}(c)), the
second-neighbor hopping benefits greatly from the phase coherence
(the same sign in Fig.~\ref{Fig:wannier}(d)) of the overlap.
Such symmetry consideration should hold for all 2H-TMDs of the same
class, due to their similar local environment around the transition
metal sites.}

\begin{figure}[t]
\includegraphics[width=3in]{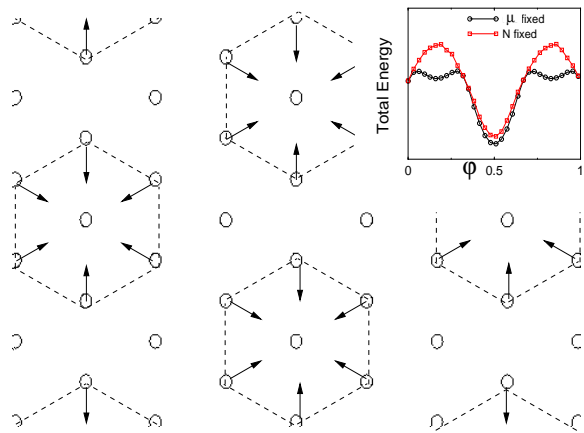}
\caption{Atomic displacement pattern corresponding to the phase
$\varphi=\pi/2$ in Eq.~(\ref{Eq:dist}). The inset shows the total
energy of the CDW state as the function of phase $\varphi$
where the total energy units are arbitrary and the
phase is in units of $\pi$.
}
\label{Fig:dis}
\end{figure}

Specifically in 2H-TaSe$_2$, this unusual electronic structure
provides a plausible intuitive resolution to the puzzling
experimental observation of gapless excitations in the CDW phase.
Indeed, with the dominating second neighbor hopping in a triangular
lattice, the system effectively splits into \emph{three} weakly
coupled sublattices. As we discuss below, one of the sublattices
remains undistorted in the CDW phase (see Fig.~\ref{Fig:dis}) and
therefore the bands associated with it are ungapped.

\begin{figure}[t]
\includegraphics[width=3in]{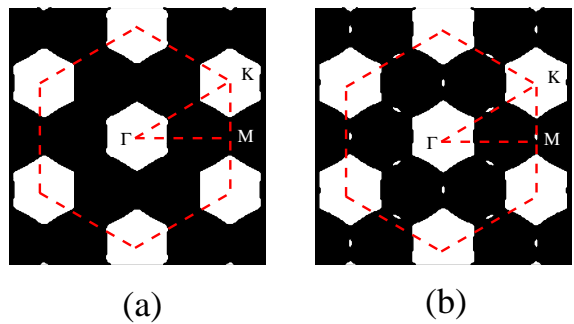}
\caption{(a) The Fermi surface from the tight-binding band
structure.  White indicates unoccupied states and black indicates
occupied states. (b) The Fermi surface for a slightly smaller
chemical potential showing the extended saddle bands \red{(small
white regions)}. The first Brillouin zone is the hexagonal cell
formed by the red \red{dashed} lines.} \label{Fig:BZ}
\end{figure}

With the dominance of the second neighbor hopping established, we
now move on to construct a minimal\red{, `nesting'} model which
captures the essential physics of \red{the gapless CDW in} 2H-TMDs.
We start with the simple \red{2D} tight-binding energies given by
\begin{equation}
\label{Eq:bs} \varepsilon_{\bf k}^{0}=\sum_{\bf R} t_{|{\bf R}|}
\cos ({\bf k} \cdot {\bf R})
\end{equation}
where $\bf R$ runs over the triangular lattice defined by lattice
vectors ${\bf a}_{1} = a (\sqrt{3}/2,1/2)$ and ${\bf a}_{2} = a
(\sqrt{3}/2,-1/2)$. \red{In addition to $t_2$,
$t_6=t_2/3$~\cite{note:t6} (all other hoppings neglected) is
introduced to produce the Fermi surface of an almost perfectly
nested ``hexagonal checkerboard'' pattern similar to the recent
ARPES data with extended saddle bands (very close to but below the
Fermi energy for extended regions along $\Gamma$K)
\cite{Liu98,Straub99,Liu00,Tonjes01}, as shown in Fig.~\ref{Fig:BZ}.
The corresponding band (with $t_2$ adjusted to 140 meV) compares
reasonably well with the first principles results (see
Fig.~\ref{Fig:band}; the two metallic bands are degenerate in the 2D
model).} As we show below, \emph{even with such a perfect Fermi
surface nesting, no gap is opened upon the formation of CDW.}

Continuing the development of our minimal model, we next consider
the CDW lattice distortions.
The detailed neutron diffraction experiments of Moncton \emph{et
al.} \cite{Moncton77} have determined that the ionic displacements
have $\Sigma_{1}$ symmetry, which corresponds to longitudinal motion
of the Ta atoms in the basal plane with amplitude given by
experiment.
However, the fitting procedure to
the measured geometric structure factors was insensitive to the
overall phase $\varphi$ of the distortions. Following this work, the
atomic displacements having $\Sigma_{1}$ symmetry corresponding to
the triple period CDW in the 2H-TMDs are given by
\begin{equation}
\label{Eq:dist}
\delta {\bf R} = \sum_{\bf Q} u \cos({\bf Q} \cdot {\bf
R}+\varphi) \hat{\bf Q}.
\end{equation}
Here $u$ is the amplitude and the sum runs over the vectors ${\bf
Q}_{1}={\bf b}_{1}/3$, ${\bf Q}_{2}={\bf b}_{2}/3$, and
 ${\bf Q}_{3}=-({\bf b}_{1}+{\bf b}_{2})/3$, where the reciprocal lattice
vectors are given by ${\bf b}_{1}=\frac{2\pi}{a}(1/\sqrt{3},1)$ and
${\bf b}_{2}=\frac{2\pi}{a}(1/\sqrt{3},-1)$.
The above atomic displacements also splits the lattice
into three independent sublattices, where one of these sublattices
does not experience displacements for any $\varphi$.
We will determine the overall phase factor $\varphi$ by minimizing
the total energy. It can be seen that the magnitude of the
displacements given by Eq.~(\ref{Eq:dist}) will not depend on
$\varphi$. Thus the elastic energy of the system will not depend on
the phase of the CDW  for this model system and our problem is
reduced to finding the phase that minimizes the energy of the
conduction band. Expanding the crystal potential to first order in
$\delta{\bf R}$ given by Eq.~(\ref{Eq:dist}) leads to the
perturbation
\begin{equation}
{\cal H}'=\sum_{\bf k, Q} \Delta_{\bf k}^{\bf Q}
c^{\dagger}_{\bf k} c_{\bf k+Q}+ {\rm h. c.}
\end{equation}
where the wave vector $\bf k$ is summed over the first Brillouin
zone.
For simplicity, we assume that the change in the
hopping parameters due to the lattice distortion is proportional to
the change in the absolute distance between neighboring atoms:
$ \delta t_{\bf R R'} \propto (\delta {\bf R} - \delta {\bf R'})
\cdot ({\bf R}-{\bf R'}).$
Then
\begin{equation}
\label{Eq:Gap} \Delta_{\bf k}^{\bf Q}=-u e^{-i \varphi} \sum_{\bf R}
\gamma^{}_{|{\bf R}|} (e^{-i {\bf Q} \cdot {\bf R}}-1) e^{-i {\bf k}
\cdot {\bf R}} \hat{{\bf Q}}\cdot\hat{{\bf R}}
\end{equation}
where the lattice vector $\bf R$ is summed over the \emph{second}
nearest neighbors to the atom at the origin and $\gamma^{}_{|{\bf
R}|=2}>0$ is the electron-phonon coupling constant in the unit of
energy/distance. Since the unit cell of the distorted lattice
contains nine sites of the original lattice, the renormalized
energies $\varepsilon_{\bf k}^{(n)}$ are given by the eigenvalues of
a $9\times 9$ matrix which is rather cumbersome and we will give its
explicit expression elsewhere.

We can now write the total energy as a function of the amplitude and
phase of the distortion as
\begin{equation}
\label{Eq:Etot} E_{\rm tot}(u,\varphi)= \int_{-\infty}^{\mu}
\varepsilon \rho(\varepsilon) d\varepsilon + E_{\rm el}(u)
\end{equation}
where $ \rho(\varepsilon) = \sum_{{\bf k}, n}
\delta(\varepsilon-\varepsilon_{\bf k}^{(n)})$ is the density of
electronic states and $E_{\rm el}(u)$ is the elastic energy which,
as we noted before, is independent of the phase of the distortion.
The $\bf k$-integration is performed by using a fine Monkhorst-Pack
mesh and corresponding weights \cite{Monkhorst76} in the irreducible
Brillouin zone of the triangular lattice. We determine the chemical
potential $\mu$ by considering two extreme cases: (i) Fixed particle
number $N$, which corresponds to an isolated metallic band at the
Fermi energy. (ii) Fixed $\mu=0$, which corresponds to significant
spectral weight from the other bands at the Fermi energy.  The
realistic situation, with two metallic bands at the Fermi energy
arising from the multi-layer structure should reside somewhere
between these two extremes. We find that the unanimous minimum for
both extreme cases occurs when $\varphi=\pi/2$. 
This minimum will become more pronounced with increased 
electron-phonon coupling constant.
In
Fig.~\ref{Fig:dis} we show the corresponding atomic displacement
pattern, which is consistent with the charge maxima seen in scanning
tunneling microscopy experiments \cite{Slough86,Coleman88}.
Furthermore, to check the robustness of this result, we have
performed the same calculation, but with $t_6$ set to zero, and have
found that the minimal total energy still occurs at $\varphi=\pi/2$.

\begin{figure}
\includegraphics[width=3in]{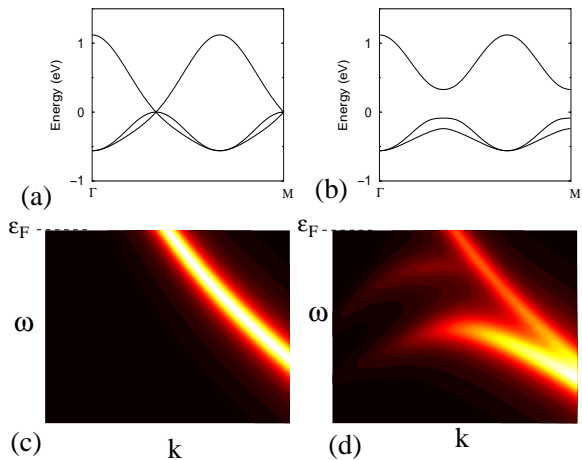}
\caption{
The bands in the low temperature CDW state originating from the undistorted
(a) and distorted (b) sublattices.  The increase in the number of bands
corresponds to backfolding resulting from the $3 \times 3$ supercell.
Theoretical ARPES spectra for the normal (c) (corresponding to vanishing
electron-phonon coupling) and CDW (d) states for wave vectors over
the nested region of the Fermi energy along $\Gamma$M.  The energy
range is $-0.47$ to $0$ eV relative to the Fermi energy and the
momentum range is $\frac{{\bf b}_{1}+{\bf b}_2}{3}\frac{3}{10}$ to
$\frac{{\bf b}_{1}+{\bf b}_2}{3}\frac{7}{10}$.
}
\label{Fig:bs}
\end{figure}

Now that the phase of the CDW has been determined we will analyze
the renormalized quasiparticle dispersion in the presence of the
CDW. The energy spectrum from the undistorted sublattice is shown in
Fig.~\ref{Fig:bs}(a) and that from distorted sublattices is shown in
Fig.~\ref{Fig:bs}(b) along the $\Gamma$M direction, which is along
the nested region of the $\Gamma$-centered hexagonal Fermi surface.
Clearly, those associated with the undistorted sublattice do not
change [Fig.~\ref{Fig:bs}(a)]. Thus, the corresponding sub-bands
remain metallic in the CDW phase. On the other hand, the bands
originating from the two distorted sublattices are doubly degenerate
and we find a gap opens at the Fermi energy [Fig.~\ref{Fig:bs}(b)].
Moreover, to make comparison with experiment more direct, in
Fig.~\ref{Fig:bs}(c) and Fig.~\ref{Fig:bs}(d) we present theoretical
ARPES spectra $ A({\bf k},\omega)=\frac{1}{\pi} f(\omega) {\rm Im}
G({\bf k},\omega)$ where $G({\bf k},\omega)$ is the single particle
Green's function and $f(\omega)$ is the Fermi distribution function
for wave vectors in a small region of the nested portion of the
Fermi surface.
To emulate experimental data, we have chosen a broadening of
$\eta=40$ meV of the spectral density function.
In Fig.~\ref{Fig:bs}(b) both the gapped and ungapped bands
are visible (for an account of the weightings of satelite bands
in CDW materials, see Voit \emph{et al.} \cite{Voit00}).
The most direct comparison between experimental data is
with the work of Valla \emph{et al.} \cite{Valla00} where there is a plot
similar to Fig.~\ref{Fig:bs}(a).  It is also shown in this
paper that no gap opens along $\Gamma$K, being
consistent with Fig.~\ref{Fig:bs}(b).

A question that naturally follows is, how robust this result is when
the finite first neighbor interaction---which mixes all bands and
thus destroys the exact decoupling into the three independent
sublattices---is taken into account. We examine this issue,
switching on the first-neighbor electron-phonon coupling constant
$\gamma_1$ in Eq.~(\ref{Eq:Gap}), up to a third of the value of
$\gamma_2$ as suggested by the first principles results: $t_1\simeq
t_2/3$. As expected, we find that the degeneracy of the bands
originating from the distorted sublattices shown in
Fig.~\ref{Fig:bs}(b) is lifted. In addition, the triple degeneracy
originating from the undistorted sublattice shown in
Fig.~\ref{Fig:bs}(a) at the Fermi energy is lifted. However, this
does $not$ produce a quasiparticle gap at the Fermi energy. More
specifically, this triple degeneracy is lifted in such a way that
two of the energies are increased to above the Fermi energy and the
other one is decreased to below the Fermi energy.  It can be seen
\ignore{through general considerations} that this will indeed not
gap the Fermi surface, given $\gamma_1$ is considerably smaller than
$\gamma_2$.

In conclusion, we have studied the CDW state in the 2H-transition
metal dichalcogenides and found that due to a unique feature in the
electronic structure of these materials revealed from
first-principles calculations, the triangular lattice can be
effectively decoupled into three independent sublattices, with one
remaining undistorted in the CDW phase. As illustrated with a model
calculation, this leads to the remarkable situation where no regions
of the entire Fermi surface become gapped even when these materials
exhibit a commensurate CDW.

The work done at Harvard University was supported by the NSF
(DMR-0132874, DMR-0231631), the Sloan Foundation, and
 Harvard NSEC and MRSEC (DMR-0213805). The work done
at Brookhaven National Laboratory was supported by U.S. Department
of Energy under Contract No. DE-AC02-98CH1-886 and DOE-CMSN. W.Y.
and W.K. thank A.M. Tsvelik and T. Valla for helpful discussions.


\begin{thebibliography}{26}
\bibitem[{not({\natexlab{a}})}]{note:wien}
\bibinfo{note}{We applied the WIEN2k \cite{wien} implementation of the full
  potential linearized augmented plane wave method in the local density
  approximation of density functional theory with the crystallographic data
  \cite{Moncton77}. The basis size was determined by $R_{mt}K_{max}=7$ and the
  Brillouin zone was sampled with a regular mesh containing 99 irreducible
  $k$-points to achieve energy convergence of 1 meV.}
\bibitem[{not({\natexlab{b}})}]{note:t6}
\bibinfo{note}{\red{$t_6$ calculated from first principles is vanishingly
small.}}
\end{thebibliography}

\end{document}